\documentstyle[prl,aps,multicol]{revtex}

\renewcommand{\narrowtext}{\begin{multicols}{2} \global\columnwidth20.5pc}
\renewcommand{\widetext}{\end{multicols} \global\columnwidth42.5pc}
\begin{document}
\title{Comment on ``Quantum key distribution without alternative measurements''}
\author{Yong-Sheng Zhang, Chuan-Feng Li\thanks{%
Electronic address: cfli@ustc.edu.cn}, Guang-Can Guo\thanks{%
Electronic address: gcguo@ustc.edu.cn}}
\address{Laboratory of Quantum Communication and Quantum Computation and Department\\
of Physics, University of Science and Technology of China, Hefei 230026, \\
People's Republic of China}
\maketitle

\begin{abstract}
\baselineskip12ptIn a recent paper [A. Cabello, Phys. Rev. A {\bf 61},
052312 (2000)], a quantum key distribution protocol based on entanglement
swapping was proposed. However, in this comment, it is shown that this
protocol is insecure if Eve use a special strategy to attack.

PACS number(s): 03.67.Dd, 03.67.Hk, 03.65.Bz
\end{abstract}

\narrowtext
\baselineskip12ptIn a recent paper \cite{Cab00}, Cabello presented a quantum
key distribution (QKD) protocol based on entanglement swapping \cite{Ben93}.
A strategy of attack by the eavesdropper (Eve) using a pair of entangled
particles was discussed, and the protocol is shown to be secure in that
case. However, here we will show that Eve can obtain the key without being
detected by the communication parties with a pair of entangled particles.

For convenience, we use the same notation as in Ref. \cite{Cab00}. The four
Bell states are denoted by 
\begin{equation}
\left| 00\right\rangle _{ij}=\frac 1{\sqrt{2}}\left( \left| 0\right\rangle
_i\otimes \left| 0\right\rangle _j+\left| 1\right\rangle _i\otimes \left|
1\right\rangle _j\right) ,  \eqnum{1}
\end{equation}
\begin{equation}
\left| 01\right\rangle _{ij}=\frac 1{\sqrt{2}}\left( \left| 0\right\rangle
_i\otimes \left| 0\right\rangle _j-\left| 1\right\rangle _i\otimes \left|
1\right\rangle _j\right) ,  \eqnum{2}
\end{equation}
\begin{equation}
\left| 10\right\rangle _{ij}=\frac 1{\sqrt{2}}\left( \left| 0\right\rangle
_i\otimes \left| 1\right\rangle _j+\left| 1\right\rangle _i\otimes \left|
0\right\rangle _j\right) ,  \eqnum{3}
\end{equation}
\begin{equation}
\left| 11\right\rangle _{ij}=\frac 1{\sqrt{2}}\left( \left| 0\right\rangle
_i\otimes \left| 1\right\rangle _j-\left| 1\right\rangle _i\otimes \left|
0\right\rangle _j\right) ,  \eqnum{4}
\end{equation}
where $i,j$ are labels of the particles.

The eavesdropping strategy is illustrated in Fig. 1 and can be described as
follows. In the beginning, Alice has particles 1 and 2 in state $\left|
11\right\rangle _{12}$, and 3 and 5 in state $\left| 10\right\rangle _{35}$.
Bob has particles 4 and 6 in state $\left| 10\right\rangle _{46}$. All this
information is public. Eve prepares particles 7 and 8 in state $\left|
10\right\rangle _{78}$.

\begin{center}
{\bf Figure 1}
\end{center}

(i) Alice sends particle 2 to Bob using a public channel and makes a Bell
type measurement on particles 1 and 3. Eve intercepts and keeps this
particle and sends her particle 7 to Bob impersonating particle 2 Alice
sends out.

(ii) Bob makes a Bell type measurement on 7 and 4, then sends particle 6 to
Alice. Eve intercepts it and makes a Bell measurement on 6 and 8.

(iii) Eve makes a unitary transformation on particle 2 according to the
measurement result of particles 6 and 8. She makes transformation $I,$ $X,$ $%
Y,$ or $Z$ corresponding to her measurement result $\left| 10\right\rangle $%
, $\left| 00\right\rangle $, $\left| 01\right\rangle $, or $\left|
11\right\rangle $ respectively. $I,$ $X,$ $Y,$ and $Z$ are operators 
\begin{eqnarray}
I &=&\left( 
\begin{array}{cc}
1 & 0 \\ 
0 & 1
\end{array}
\right) ,\text{ }X=\left( 
\begin{array}{cc}
0 & 1 \\ 
1 & 0
\end{array}
\right) ,\text{ }  \eqnum{5} \\
Y &=&\left( 
\begin{array}{cc}
0 & -1 \\ 
1 & 0
\end{array}
\right) ,\text{ }Z=\left( 
\begin{array}{cc}
1 & 0 \\ 
0 & -1
\end{array}
\right) .
\end{eqnarray}
Then Eve sends particle 2 to Alice impersonating the particle 6 Bob sends
out.

(iv) Alice makes a Bell type measurement on 5 and 2 and publicly announce
the result. Thus Alice's and Bob's results of measurement will be consistent
as if there is no eavesdropper intervening.

The reason that Eve makes a unitary operation on particle 2 is as follows.
Assume that Alice's particles 2 and 5 are in state $\left| \Phi
\right\rangle $, if Eve does not intervene and Bob's result of Bell type
measurement is $\sigma \left| 10\right\rangle $ ($\sigma $ is one of $I$, $X$%
, $Y$ and $Z$), Alice's result of Bell type measurement will be $\sigma
\left| \Phi \right\rangle $. Now, Eve intervene the process, her result of
the Bell measurement on 6 and 8 will be the same as Bob's result of the Bell
measurement on 7 and 4. For the consistent of Alice's and Bob's measurement,
if Eve obtain the result $\sigma \left| 10\right\rangle $, she ought to
makes a transformation $\sigma $ on particle 2, so that when Alice measures
particles 2 and 5, she obtains the proper result $\sigma \left| \Phi
\right\rangle $.

For example, suppose that Alice obtains ``11'' in her measurement on
particles 1 and 3, and Alice can know that the state of 5 and 2 is $\left|
10\right\rangle _{25}$. Suppose that Bob has obtained ``00'' in his
measurement on 7 and 4. Eve will obtain ``00'' in her measurement on 6 and 8
too, then she makes a transformation $X$ on particle 2 and sends it to
Alice. Alice makes a Bell measurement on 2 and 5 and will obtain the result
``00''. From Table I in \cite{Cab00}, Alice knows that Bob has obtained
``00'' and Bob can know Alice's initial result is ``11''. Alice's result
``11'' will be the key bits between them. Because Eve knows Bob's result and
Alice's public announcement of the measurement result of 2 and 5, she can
know Alice's initial result ``11'' from Table I too.

All steps above will introduce no error in the key distribution between
Alice and Bob, and Eve can know exactly the result of Bob's measurement in
step (ii) and also the public information publicly announced in step (ii).
So Eve can obtain the key they distributed successfully. In conclusion, this
protocol is insecure against this type attack.

This work was supported by the National Natural Science Foundation of China.

{\bf Figure caption:}

{\bf Figure1}. Eve's strategy to obtain Alice's secret result. The notations
are the same as in Ref. \cite{Cab00}. The bold lines connect qubits in Bell
states, the dashed lines connect qubits on which a Bell operator measurement
is made, and the pointed lines connect qubits in Bell states induced by
entanglement swapping. ``00'' means that the bell state $\left|
00\right\rangle $ is public knowledge, (00) means that it is only known to
Alice, [00] means that it is only known to Bob, 
\mbox{$\vert$}
00%
\mbox{$\vert$}
means that it is unknown to all the parts, \{00\} means that it is only
known to Eve, [(00)] means that it is known to Alice and Bob (and Eve), etc.

\widetext

\end{document}